\documentclass[twocolumn]{webofc}

\usepackage[varg]{txfonts}   
\usepackage{hyperref}
\usepackage{url}
\usepackage{setspace}
\usepackage{physics}
\usepackage{xcolor}

\hypersetup{colorlinks=true,citecolor=blue,urlcolor=blue,linkcolor=blue}

\begin{document}

\title{Vortex Creep Heating in Neutron Star Cooling: New Insights into Thermal Evolution of Heavy Neutron Stars}

\author{
  \firstname{Yoonhak} \lastname{Nam} \inst{1}\fnsep\thanks{\email{nam.y.b76c@m.isct.ac.jp}} \and
  \firstname{Kazuyuki} \lastname{Sekizawa} \inst{1,2,3}\fnsep\thanks{\email{sekizawa@phys.sci.isct.ac.jp}}
}

\institute{
  Department of Physics, School of Science, Institute of Science Tokyo, Tokyo 152-8550, Japan
  \and
  Nuclear Physics Division, Center for Computational Sciences, University of Tsukuba, Ibaraki 305-8577, Japan
  \and
  RIKEN Nishina Center, Saitama 351-0198, Japan
}

\abstract{Neutron stars provide unique laboratories for probing physics of dense nuclear matter under extreme conditions. Their thermal and luminosity evolution reflects key internal properties such as the equation of state (EoS), nucleon superfluidity and superconductivity, envelope composition, and magnetic field, and so on. Recent observations [\textit{e.g.}, V. Abramkin \textit{et al.,} ApJ \textbf{924}, 128 (2022)]  have revealed unexpectedly warm old neutron stars, which cannot be explained by standard neutrino–photon cooling models. The failure of the standard cooling models implies the presence of additional internal heating mechanism. Building on the previous study [M. Fujiwara \textit{et al}., JCAP \textbf{03}, 051 (2024)], which proposed vortex creep heating (VCH) from the frictional motion of superfluid vortices as a viable mechanism, we extend the cooling framework to include both VCH and direct Urca (DUrca) processes. These are implemented in our code to explore their combined impact, particularly for massive neutron stars where DUrca operates. By varying rotational parameters ($P$, $\dot{P}$, $P_0$), EoS models (APR, BSk24), pairing gaps, and envelope compositions, we examine how heating–cooling interplay shapes the temperature evolution. Our results show that VCH can substantially mitigate the rapid cooling driven by DUrca, offering new evolutionary pathways for massive neutron stars.
}
\maketitle
\vspace{-5mm}

\section{Introduction}
\label{intro}

Neutron stars are unique natural laboratories for studying nuclear matter under extreme conditions. 
Their thermal and luminosity evolution sensitively reflects key internal properties, including the dense-matter equation of state (EoS), nucleon superfluidity and superconductivity, envelope composition, and magnetic field.

Because neutron stars rotate, their internal neutron superfluid forms quantized vortices that interact with the Coulomb lattice of nuclei in the inner crust, influencing both rotational dynamics (\textit{e.g.}, glitches) and thermal evolution. When vortices unpin and migrate outward---a process known as \textit{vortex creep}---frictional dissipation occurs, producing internal heating. This mechanism, referred to as \textit{vortex creep heating} (VCH), has been proposed to explain the unexpectedly high temperatures of old neutron stars (\textit{e.g.}, Ref.~\cite{Abramkin_2022_recent_observation_7}) that cannot be accounted for by standard neutrino and photon cooling alone.

Most previous studies (\textit{e.g.}, Ref.~\cite{Fujiwara_2024}) examined VCH in low-mass stars where direct Urca (DUrca) processes are forbidden, while its role in massive stars allowing DUrca reactions remains largely unexplored. 
In this work, we have implemented the VCH to our newly-developed neutron-star cooling code \cite{Nam_2025_data_driven_cas_a_ns}, and further extended the framework to include both vortex creep heating and DUrca cooling. By varying rotational parameters ($P$, $\dot{P}$, $P_0$), EoS models (APR \cite{Akmal_1998_APR_EoS}, BSk24 \cite{Pearson_2018_BSk24_EoS}), superfluid pairing gaps, and envelope compositions, we investigate how the interplay between heating and cooling shapes the temperature evolution across different stellar masses. In this contribution, we outline our ongoing study to explore VCH and DUrca processes in massive neutron stars, underlining a necessity of a new point of view (Fig.~\ref{fig-2}) when assessing cooling curves with VCH. A full paper is in preparation and will be published elsewhere.

\section{Methods}

\subsection{Vortex creep heating}\label{Subsec:VCH}

Superfluid vortices in the inner crust of rotating neutron stars can convert rotational energy into heat through frictional motion. In the two–component model \cite{Alpar_1984_VCH_1,Shibazaki_1989_VCH_2,Link_1996_VCH_3}, the rigidly rotating crust (ions and charged fluids) is coupled to a neutron superfluid via pinned vortex lines. An external torque decelerates the crust, while the superfluid responds through vortex creep.

\begin{figure*}[t]
\centering
\includegraphics[width=13cm,clip]{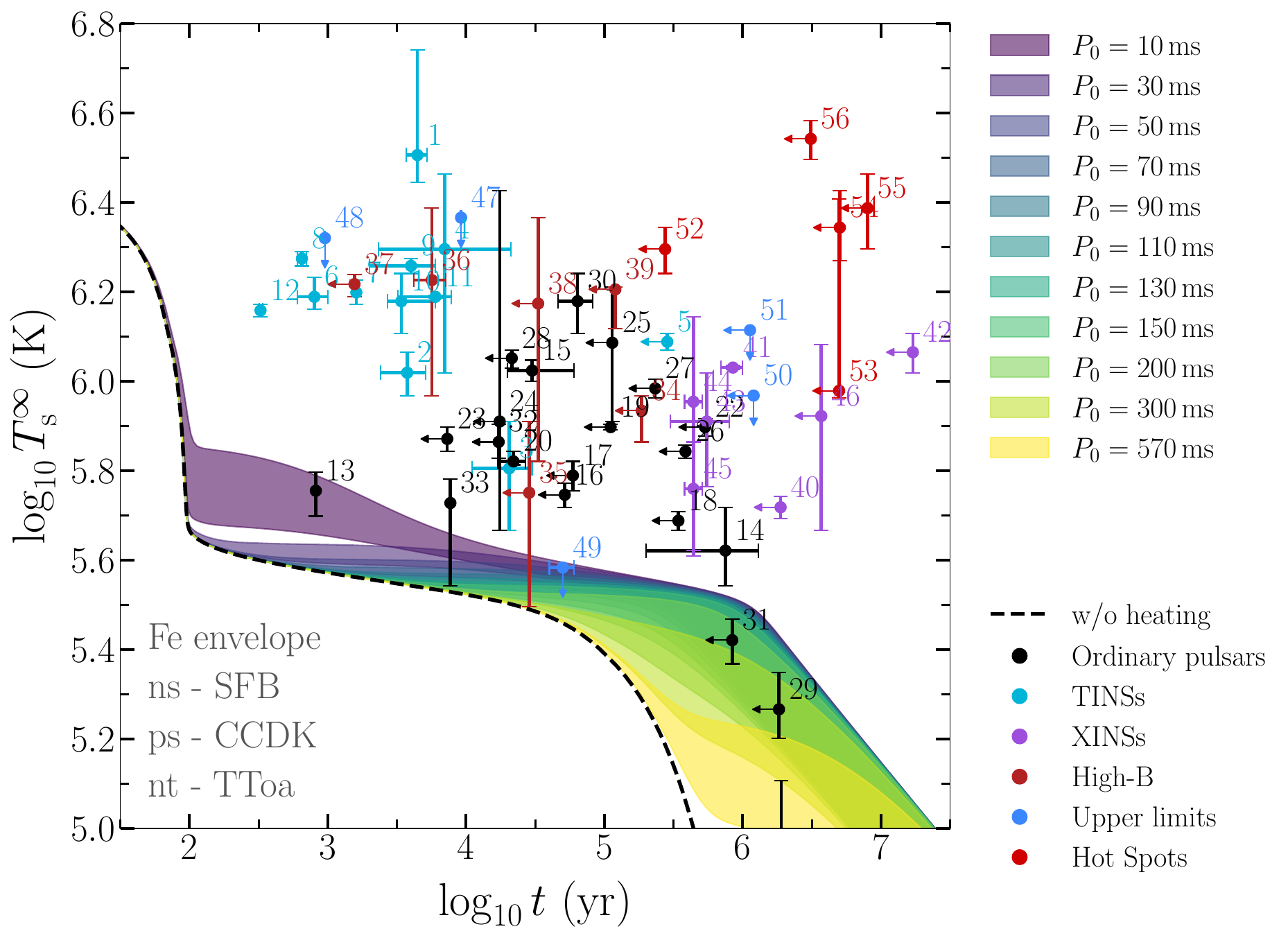}\vspace{-2mm}
\caption{
Cooling curves of $2.0\,M_\odot$ neutron stars at $B$\,$=$\,$10^{12}\,$G (BSk24; Fe envelope; ns–SFB, ps–CCDK, nt–TToa). For each $P_0$\,$=$\,10--570\,ms, the shaded band spans $J$\,$=$\,$10^{42.9}$--$10^{43.8}\,{\rm erg\,s}$; the dashed line is standard cooling without heating. Symbols show $T_\mathrm{s}^\infty$ vs.\ $t$ ($\log_{10}$ axes); colors denote classes. Left arrows mark characteristic-age estimates only; numbers label sources.}
\label{fig-1}
\end{figure*}

The angular-velocity lag, $\delta\Omega=\Omega_\text{s}-\Omega_\text{c}$, produces Magnus force per unit vortex length,
\begin{align}
\mathbf{f}_{\rm Mag}=n_\text{s}(\delta\boldsymbol{\Omega}\times\mathbf{r})\times\boldsymbol{\kappa},
\end{align}
where $n_\text{s}$ is the superfluid density, $\mathbf{r}$ the position vector from the rotation axis, and $\boldsymbol{\kappa}$ the circulation quanta ($|\boldsymbol{\kappa}|=h/2m_\text{n}$). The Magnus force is balanced by the pinning force $f_{\rm pin}$, defining the critical lag, $\delta\Omega_{\rm cr}\simeq f_{\rm pin}/(n_\text{s}\kappa r)$. For $T\gg T_q$ ($T_q\!\sim\!10^8$~K) creep is thermal, while for $T\ll T_q$ it proceeds via quantum tunneling \cite{Baym_1992_thermal_or_quantum_1, Link_1993_thermal_or_quantum_2}.  

When the creep rate is large enough to match the global spin-down, the steady-state heating luminosity becomes
\begin{align}
\nonumber\\[-7.5mm]L_\mathrm{h} = J|\dot{\Omega}_\infty|, \qquad 
J=\int_{\rm pin}\delta\Omega_{\infty}\,\dd I_\text{pin},
\end{align}
where $\dd I_\text{pin}$ is the differential moment of inertia of the pinned region \cite{Fujiwara_2024}. This long-term internal heat source counteracts cooling in old neutron stars.

\subsection{Neutron--star cooling}\label{cooling}

The thermal evolution obeys the general-relativistic energy-balance and transport equations:
\begin{align}
\frac{\dd(L e^{2\Phi})}{\dd r}
  &=-\frac{4\pi r^2 e^{\Phi}}{\sqrt{1-2Gm/c^2r}}
   \!\left(C_V\frac{\dd T}{\dd t}+e^{\Phi}(Q_\nu-Q_\mathrm{h})\right),\\[-1mm]
\frac{\dd(Te^{\Phi})}{\dd r}
  &=-\frac{L e^{\Phi}}{4\pi r^2\lambda\sqrt{1-2Gm/c^2r}},
\end{align}
where the heating rate $Q_\mathrm{h}$ includes VCH. After the core becomes isothermal, they are reduced to the global form
\begin{align}
L_\gamma^\infty = -C\,\frac{\dd T_\mathrm{b}^\infty}{\dd t} - L_\nu^\infty + L_\mathrm{h}^\infty.
\end{align}
We solve these equations with our newly-developed neutron-star cooling code written in Fortran90 \cite{Nam_2025_data_driven_cas_a_ns}, which follows the numerical framework of \texttt{NSCool} \cite{Page_2016_NSCool}. The stellar structure $(M,R,\Phi)$ is obtained from the TOV equations using the APR \cite{Akmal_1998_APR_EoS} and BSk24 \cite{Pearson_2018_BSk24_EoS} EoSs. Neutrino processes include modified and direct Urca and Cooper-pair breaking and formation (PBF), with pairing gaps from the CCDK \cite{Elgaroy_1996_ccdk} (for $^1\mathrm{S}_0$ protons), SFB \cite{Schwenk_2003_SFB} (for $^1\mathrm{S}_0$ neutrons), and TToa \cite{Takatsuka_2004_nt_TTav_TToa} (for $^3\mathrm{P}_2$ neutrons) models.

\begin{figure*}[t]\vspace{-2mm}
\centering
\includegraphics[width=16cm,clip]{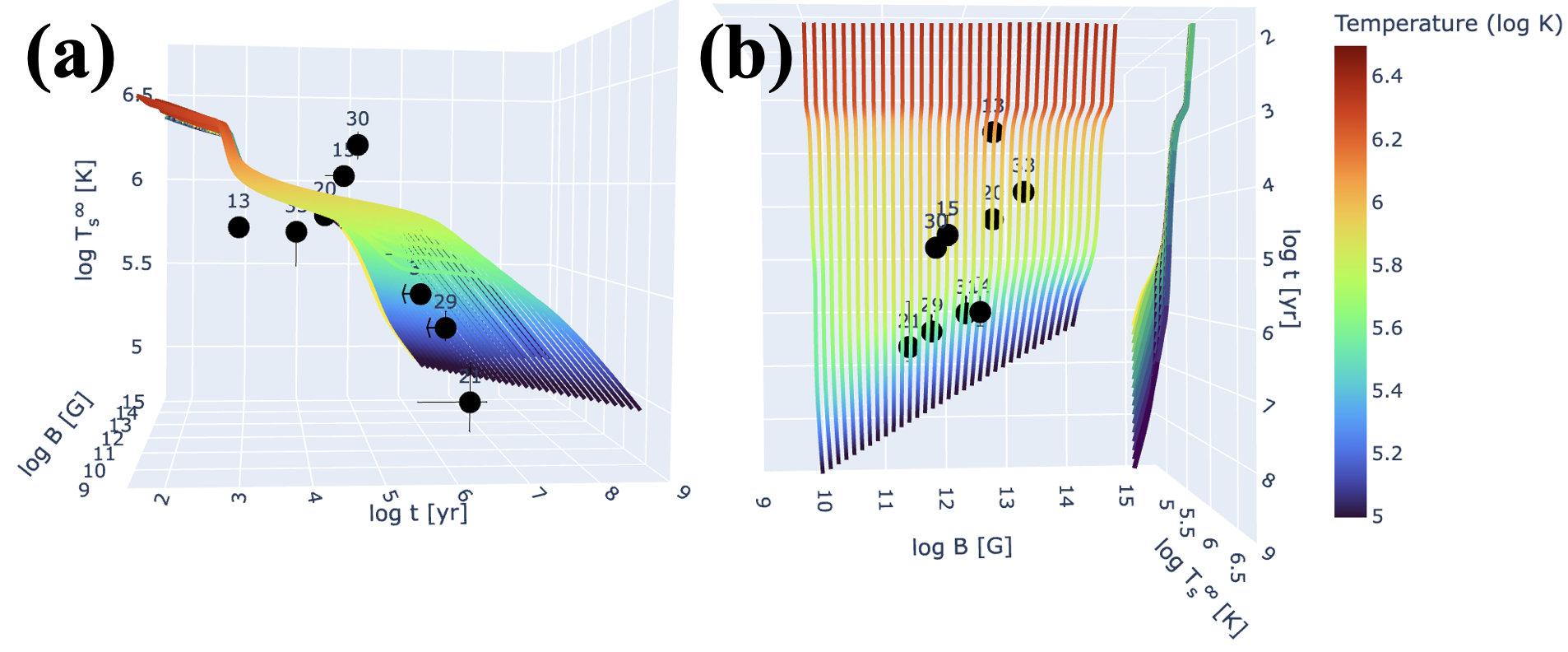}\vspace{-3mm}
\caption{
Cooling evolutions of $1.4\,M_\odot$ neutron stars at fixed $P_0$\,$=$\,10\,ms and $J$\,$=$\,$10^{43.8}$\,{erg\,s} for different magnetic-field strengths $B$. (a)~3D representation of the cooling surface, where the conventional $(t,\,T_\mathrm{s}^\infty)$ cooling curves are extended along the magnetic-field axis. (b)~Projection along the temperature axis, showing the time evolution at different $B$ values, with color indicating $\log_{10} T_\mathrm{s}^\infty$. Black circles denote observed ordinary pulsars with measured surface temperatures and estimated magnetic fields.
}\vspace{-1mm}
\label{fig-2}
\end{figure*}

\subsection{Observational data}\label{obda}

We use observational data compiled in the Ioffe Institute neutron-star database maintained by Potekhin and collaborators \cite{Potekhin_data_of_ioslated_ns}. This database provides a homogeneous and regularly updated collection of neutron-star thermal-emission parameters, including the spin period $P$, its time derivative $\dot{P}$, dipolar magnetic field $B_{\mathrm{dip}}$, characteristic and kinematic ages ($t_\textit{c}$ and $t_*$, respectively), and the redshifted surface temperature $T_\infty$. Sources are categorized by magnetic-field strength and emission properties into weakly magnetized thermal emitters, ordinary pulsars, high-$B$ pulsars, and the ``Magnificent Seven.'' These data serve as the observational basis for comparison with our theoretical cooling curves.

\section{Results and Discussion}

In Fig.~\ref{fig-1}, we show calculated cooling curves together with observational data. From the figure, we find that faster initial rotation (smaller $P_0$) enhances early VCH, sustaining higher surface temperatures during $10^3$–$10^5$~yr. The width of the shaded bands reflects uncertainty in the proportional constant $J$. Once the system enters the quantum–creep regime ($T\!\ll\!T_q$), the heating power becomes nearly temperature–independent, and for larger $P_0$, the onset of heating occurs later in the evolution.

We underline here the necessity of extending the plot to incorporate $B$ dependence. Namely, in Fig.~\ref{fig-2}, a new $\log_{10}B$-axis has been added to the usual $\log_{10}t$--$\log_{10}T_\text{s}^\infty$ plot, revealing distinct $B$ values. It indicates that stronger magnetic fields cause slightly faster late-time cooling. Although the $B$ dependence alone is weaker than that on $P_0$, its combined effect with the initial spin period can still influence the onset and duration of VCH.

It is worth noting that in Fig.~\ref{fig-1}, the $P_0$\,$=$\,10\,ms curve appears to pass through data points~13, 29, 31, and~33. However, as Fig.~\ref{fig-2} clarifies, these sources possess different magnetic-field strengths. Therefore, even for the same $P_0$, the corresponding cooling curves deviate depending on $B$, highlighting the importance of treating the magnetic field as an independent parameter in the thermal evolution if VCH is considered.

Overall, incorporating both VCH and the DUrca process, we have revealed that old, thermally bright neutron stars---previously difficult to explain by standard cooling models---could in fact be more massive objects whose rapid neutrino cooling is partially compensated by internal heating. This suggests that accounting for the interplay between heating and fast neutrino emission is essential for interpreting the thermal states of old neutron stars.

\section{Summary and Conclusion}

In summary, we have further extended our brand-new neutron-star cooling code \cite{Nam_2025_data_driven_cas_a_ns} to incorporate the vortex-creep heating (VCH). In this contribution, we have reported an example of applications for a $2M_\odot$ neutron star, where direct Urca (DUrca) fast neutrino emissions play a significant role. Such an interplay between VCH and DUrca processes was not explored well in the preceding studies. From the results, we have found that VCH causes a sizable effect on cooling curves after activation of DUrca rapid-cooling processes, implying a possibility of presence of warm old neutron stars which are much more massive than previously thought.

Moreover, we have advocated the necessity of introducing 3D plots (\text{cf.} Fig.~\ref{fig-2}) that incorporate not only the conventional cooling curve $(t, T^\infty_{\mathrm{s}})$ but also the magnetic field strength $B$ as an additional axis, if the VCH scenario is indeed valid. This novel visualization enables a clearer understanding of how neutron stars undergo distinct thermal evolution depending on the magnitude of their magnetic field.

Looking ahead, we will extend our computational framework to highly magnetized neutron stars by implementing a two-dimensional cooling solver that treats anisotropic thermal conduction in strong magnetic fields. We further plan to calibrate the proportional constant $J$ via microscopic calculations of the pinning force $f_{\mathrm{pin}}$ under multiple equation of state (EoS) models (see, \textit{e.g.}, Refs.~\cite{Wlazlowski(2016),Klausner(2023)}, enabling exploration of $J$-EoS dependencies. Finally, we intend to incorporate internal heating mechanisms other than VCH, such as rotochemical heating (see, \textit{e.g.}, Refs.~\cite{Reisenegger_1995_rotochemical_1, Fernandez_2005_rotochemical_2, Petrovich_2010_rotochemical_3}), magnetic field decay (see, \textit{e.g.}, Refs.~\cite{Pons_2009_magnetic_field_decay_1, Vigano_2013_magnetic_field_decay_2}), and dark-matter heating (see, Refs.~\textit{e.g.}, \cite{Kouvaris_2008_dark_matter_1, Lavallaz_2010_dark_matter_2, Baryakhtar_2017_dark_matter_3}).

\vspace{-1mm}
\section*{Acknowledgments}\vspace{-0.5mm}

This work is supported by JSPS Grant-in-Aid for Scientific Research, Grants No.~23K03410, No.~23K25864, and No.~JP25H01269.

\vspace{-2mm}
\bibliography{ref}

\end{document}